# Tailoring the Dimensionality of Tellurium Nanostructures via Vapor Transport Growth

Sara Ghomi[a,b], Pinaka Pani Tummala[a,c,d], Raimondo Cecchini[e], Carlo S. Casari[b], Alessio Lamperti[a], Carlo Grazianetti[a], Christian Martella[a,*], and Alessandro Molle[a]

[a] *CNR-IMM, Unit of Agrate Brianza, via C. Olivetti 2, I-20864, Agrate Brianza Italy*
[b] *Dipartimento di Energia, Politecnico di Milano, via Ponzio 34/3, Milano 20133, Italy*
[c] *Department of Physics and Astronomy, KU Leuven, Celestijnenlaan 200D, Leuven 3001, Belgium*
[d] *Dipartimento di Matematica e Fisica, Università Cattolica del Sacro Cuore, via della Garzetta 48, Brescia 25133, Italy*
[e] *CNR-IMM, Unit of Bologna, via Gobetti 101, Bologna, 40129, Italy*

**Abstract**

The interest in tellurium nanostructures is on the rise due to their outstanding physical properties including high carrier mobility, anisotropic charge conduction, photoconductivity, thermoelectricity, and piezoelectricity. Applications in related technologies require tailoring the synthesis of tellurium from its preferred vertical growth toward the lateral growth. Here, the synthesis of pillar-like and pennette-like structures of tellurium through a powder vapor deposition technique has been discussed. It has been shown that exploiting salt additives such as NaCl in vapor deposition technique can enhance tellurium pillar dimensionality toward large planar grains. Further, we report on the synthesis of hexagonal ultrathin tellurium nanoflakes, namely tellurene, from few-layer to monolayer thickness by optimizing the growth kinetics without the usage of any additives. In addition, we explore the surface quality and physical properties of as-grown two-dimensional (2D) tellurium, using a variety of characterization techniques, including Raman spectroscopy, scanning electron microscopy, atomic force microscopy, and Kelvin probe force microscopy. This study provides a pivotal scheme for enabling scalable 2D tellurium integration in numerous potential applications for electronics and optoelectronic devices.

## Introduction

In tandem with the emergence of ultrathin-layered materials, two-dimensional (2D) monoelemental Xenes such as tellurene, are on the rise. Tellurene is found to overcome fundamental limitations of other 2D materials such as the zero-bandgap of graphene, low charge carrier mobility of $MoS_2$, and ambient instability of black phosphorus [1–3]. On the other hand, tellurene poses superior properties such as photoconductivity, thermoelectricity, and piezoelectricity which make it an intriguing candidate for modern technological applications [4–7]. As a group-VI semiconducting chalcogen, tellurium (Te) is composed of covalently bonded atoms in one-dimension (1D) chiral chains along the c-axis bound by weak van der Waals forces arranged in a hexagonal lattice, which contributes to Te anisotropic atomic structure and its tendency to grow in a unidirectional manner along the c-axis, forming various 1D structures [8–12]. Te growth in 1D has been studied extensively and prepared using different methods such as hydrothermal, solution-based synthesis, thermal evaporation, low temperature, and high temperature vapor phase depositions [9,13–18]. However, to unlock the potential of Te for technological applications, developing growth methods that enable large-area growth of 2D Te is crucial. To date, the growth of 2D Te is mostly based on solution-based methods [19–22], and physical vapor deposition [23–28]. However, the utilization of salt-made additives like alkali metal halides in vapor deposition methods and their influence on nucleation and morphological evolution of Te have remained largely unexplored. Salt-made additives have been found to be effective in controlling the synthesis and the structural properties of popular 2D materials, such as transition metal dichalcogenides, via chemical vapor deposition (CVD) methodologies [29,30]. This includes increasing reaction rates, lowering melting points, and decreasing activation energies [31]. Yet, no reports on their application to the vapor deposition growth of Te have been published. In addition to the surface scalability, controlling Te thickness by adjusting its growth kinetics is crucial as well. Although there are limited reports on sub-nm thin tellurene growth, including monolayer and few-layer growth in molecular beam epitaxy [28] and tri-layer growth in vapor deposition techniques [24], the ability to thin down to monolayer hexagonal flakes by tuning the growth conditions remains scarcely explored. In this work, we employ two simple routes to master the dimensionality of Te nanostructures. First, by implementing a facile salt-assisted vapor transport method, the inherited vertical growth of Te will be suppressed, paving the way to enhance its lateral growth over the cm$^2$ area. Second, tailoring the growth kinetics in a single step vapor deposition technique is developed to achieve hexagonal flakes of tellurene with thicknesses ranging from 0.4 nm to few-nm and lateral sizes varying between 4 to 10 μm.

---

* Corresponding author.
  E-mail address: christian.martella@mdm.imm.cnr.it

**Experimental**

Synthesis

In the vapor deposition reactor consisting of a 2" quartz tube, 20 mg Te powder in an alumina boat was placed at the center of the upstream furnace. In the first set of experiments the SiO$_2$/Si serving as the substrate was placed onto the ceramic boat, at different positions within the tube: (i) position 1: 25 cm upstream of the Te powder (ii) position 2: 15 cm upstream of Te powder and (iii) position 3: 35 cm downstream of Te powder as illustrated schematically in **Figure 1a**. The temperatures of 650 °C for the upstream and 625 °C for the downstream furnace are set respectively, with a 100 sccm Ar/H$_2$ flux (H$_2$ 4% volume) of carrier gas. The second set of experiments was run with 20 mg Te powder inside the alumina boat at the center of upstream furnace with temperature of 440 °C, and a piece of SiO$_2$/Si substrate located on a ceramic boat at the center of downstream furnace with temperature of 350 °C and a flux of 10 sccm Ar/H$_2$ was used for the carrier gas within the 30 minutes of growth time.

Characterization

The morphology of the samples was examined using a Zeiss-SUPRA 40 field-emission SEM equipment in bright-field mode. The growth of Te was checked by Raman spectroscopy (Renishaw InVia spectrometer) equipped with a solid-state laser source of excitation wavelength 514 nm / 2.41 eV. Atomic force microscope (AFM, Bruker Dimension Edge) equipped with ultrasharp silicon probes (nominal tip radius < 10 nm), in the tapping mode, was employed to obtain the morphology of the as-grown Te. Kelvin probe measurements were conducted to probe the surface potential in ambient at room temperature using a conductive Pt/Ir AFM-tip and electrically grounding the sample surface by means of the microscope sample holder.

Numerical simulation

The 2D macroscale model of the double-furnace reactor was developed using the finite element software COMSOL Multiphysics. As part of this model, Navier-Stokes equations, mass diffusion, and conduction and convection heat transfer are modeled using finite element method (FEM) applying appropriate initial and boundary conditions. Based on the simulation, the velocity of the carrier gas, precursor flux distribution, and temperature distribution are predicted. In details the Navier-Stokes equation governing the laminar flow is:

$$\rho \left(\frac{\partial u}{\partial t} + u \cdot \nabla u\right) = -\nabla p + \nabla \cdot \left(\mu(\nabla u + (\nabla u)^T) - \frac{2}{3}\mu(\nabla \cdot u)I\right) + F; \quad \frac{\partial \rho}{\partial t} + \nabla \cdot (\rho u) = 0 \qquad (1)$$

where u is the velocity field, ρ is the mass density, μ the dynamic viscosity, p the pressure, I the unit matrix, and F the volumetric applied force (i.e., gravity). The superscript T means transposed matrix.

The equation describing the mass transport is:

$$R = \frac{\partial c}{\partial t} + \nabla \cdot (-D\nabla c) + u \cdot \nabla c; \quad N = -D\nabla c + u \qquad (2)$$

where R is the source term for precursor, c is the precursor concentration, D is the diffusion coefficient, and N is the flux of the precursor.

The conduction and convection heat transfer are described by:

$$\rho C_p = \frac{\partial T}{\partial t} + \rho C_p \mathbf{v} \cdot \nabla T + \nabla \cdot \mathbf{q} =; \mathbf{q} = -k \nabla T \tag{3}$$

Where the $C_p$ is the specific heat of gas, the k is thermal conductivity and T is temperature.

## Results and discussion

### 1.1. Position-dependent growth

The schematic of the upstream and downstream furnaces with Te precursor boats and the SiO$_2$/Si substrates are depicted in **Figure 1a**. Applying the temperature ramp shown in **Figure SI-1a** for the upstream and downstream furnaces, the Te vapors deposit on SiO$_2$/Si substrates located at position 1, 2, and 3. Temperatures upstream and downstream are maintained at 650 ºC and 625 ºC, respectively. Different Te nanostructures grow along the quartz tube at different positions. **Figure 1b-e** summarizes the SEM analysis performed on Te deposited on SiO$_2$/Si substrate. Under the growth conditions dictated at each position of substrates 1 and 2 such as temperature and the distance from the Te precursor, the observed morphology of the as-grown Te is different. While Te deposited on substrate 1 shows a dense solid-pillar like structure, Te deposited on substrate 2 has a layered-pillar growth normal to the substrate. The average diameter distribution of the as-grown Te solid-pillar and layered-pillar structures on substrate 1 and substrate 2 are discussed in **Figure SI-2**. Although substrates 1 and 2 are located upstream of the precursor powder, Te is still being deposited, which we will discuss in more detail below. In the case of substrate located at position 3, the morphology of the as-grown Te is different and shows "pennette-like" Te nanotubes (**Figure SI-3**), which are empty inside and with random orientation distributed over the substrate (**Figure 1e**). Due to the relatively long distance of this substrate from the Te precursor source, the Te flux toward the substrate is much lower and the temperature is expected to be higher than substrates 1 and 2 (see simulation results in the following), which in turn can result in the evolution from 1D nanostructures to tube-like Te structure ("pennette-like") [32].

The large-scale SEM images of Te nanostructures grown on substrate 1, substrate 2 without NaCl promoter, substrate 2 with NaCl promoter, and substrate 3 are presented in **Figure SI-4** confirming the distribution of nanostructures on a relatively large scale. In addition, the homogeneity level of Te nanostructures grown on each of the substrates are mapped by means of positional Raman spectroscopy scanning and reparented in **Figure SI-5**.

These three different positions, located in the cold area of the furnace quartz tube, show the tendency of Te to be grown in one dimension, rather than expanding over two dimensions. This fashion can be attributed to several reasons such as Te unique crystal structure composing of Te helical 1D chains arranged in a hexagonal lattice [2]. In addition, lower temperatures trigger the densification of nucleation points, resulting in high-density Te pillars closely-packed covering the full area of the substrate (2 cm × 2 cm). A further reason for Te atoms tendency to grow longitudinally is that they cannot overcome the high formation energy to form kinks at lower temperatures [33]. Considering this, we attempt to perform the growth at the same condition of position 2 (at low temperature) and use the solution of salt additive NaCl

(250 mg) in deionized (DI) water (200 ml) spin-coated homogeneously over the SiO$_2$/Si substrate and compare it with the growth results on SiO$_2$/Si located at position 2 without the usage of any salt additives.

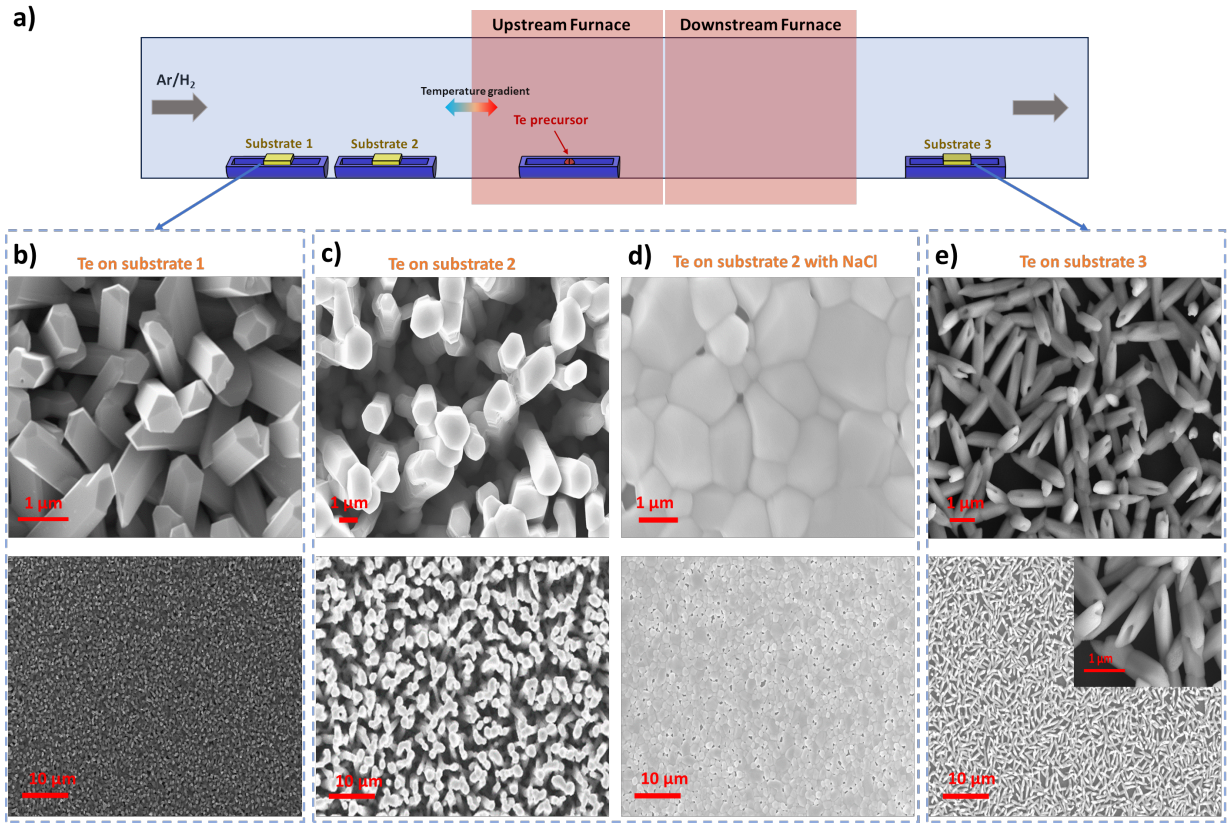

*Figure 1 a) schematic illustration of the vapor deposition experimental setup with the positions of the Te powder precursors and the SiO$_2$/Si substrates 1, 2, and 3. SEM morphological analysis of the as-grown Te nanostructures grown on SiO$_2$/Si substrates located at b) position 1 c) position 2 d) position 2 with salt spin-coated on top of the substrate e) position 3 with high magnification on top row and the corresponding SEM images with the low magnification on bottom row*

The effect of spin-coated NaCl solution in DI water on the SiO$_2$/Si substrate at position 2 was studied to determine if it promoted the lateral growth rather than vertical growth of Te. We prepared two SiO$_2$/Si substrates: one with NaCl solution spin-coated on top and the other without any additives utilized in the same growth experiment to have a more accurate comparison. The SEM morphology comparison of the Te deposited on SiO$_2$/Si and NaCl/SiO$_2$/Si are summarized in **Figure 1c-d**. It clearly shows that Te pillars with hexagonal cross-sections formed on SiO$_2$/Si substrates without NaCl (**Figure 1c**) have evolved to more tightly packed laterally grown Te on spin coated NaCl on SiO$_2$/Si substrate (**Figure 1d**).

A simulation based on finite element method was conducted to gain insight into the details of Te deposition on substrates 1 and 2 (details in experimental section). An exact 2D model of the deposition tube with substrates and precursor boats is built in this simulation. The temperature of 650 °C and 625 °C are applied to the upstream and downstream (same as experimental growth condition). The result of the simulation reveals the macroscale heat (**Figure 2a**) and evaporated Te precursor flux (**Figure 2b**) distribution over the deposition tube. The simulation predicts that there is a pronounced temperature gradient between precursor in the upstream furnace and substrates 1 and 2 outside the furnace (**Figure 2a**). The background color in **Figure 2b** and a closer view of substrates 1 and 2 in **Figure 2c** illustrate the variation of total precursor flux (mol/m$^2$.s), with lighter and darker blues representing higher and lower total flux values. As it can be observed in **Figure 2b**, although the majority of the total precursor flux is toward the downstream, in the carrier gas flow direction, a backward flow can be observed close to the substrates 1 and 2 regions. The vectors in **Figure 2c** show the direction of the total precursor flow variation. In this figure, the total precursor flux vectors in the core of the tube are toward the downstream, while the vectors at the boundary of the tube head toward the backward direction. This observation can be explained by two types of transport mechanisms that play a role in this backward deposition: convection flux in the direction of the carrier gas flow and diffusive flux opposite of the carrier gas flow toward the substrate 1 and 2 regions. As a result of this spatial variation of temperature and diffusion-driven transport, the Te particles can deviate from the actual carrier gas path, driving the precursor particles towards the substrates 1 and 2.

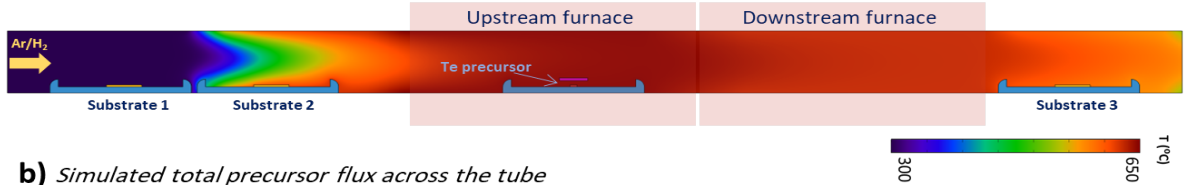

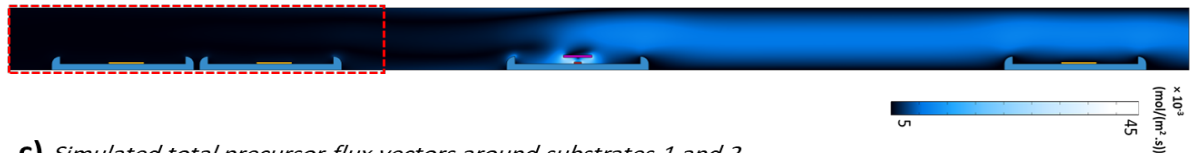

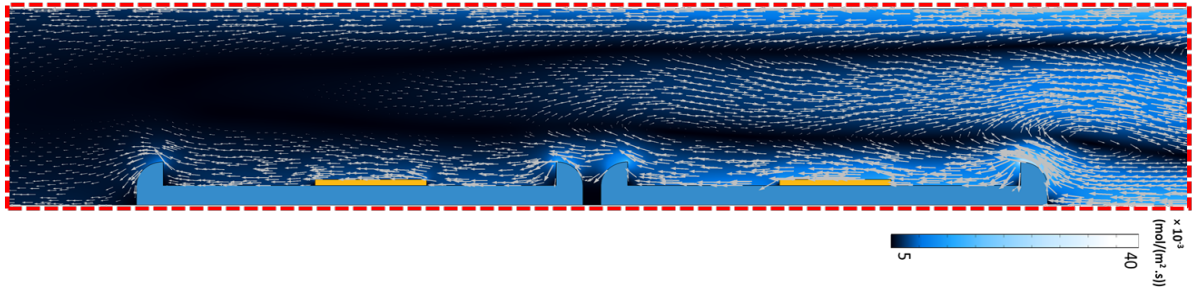

*Figure 2 FEM simulation results of the a) Temperature distribution across the double-furnace deposition system with the position of the substrates 1, 2, and 3 and Te precursor boats b) Total precursor flux variations across the double-furnace deposition system c) Total precursor flux at the substrates 1 and 2 area (background color represents the total precursor flux and the vectors height and directions are proportional to the magnitude and direction of the total flux).*

Raman spectroscopy performed on the Te deposited on substrate without NaCl located at position 2 is shown in **Figure 3a** top panel and reveal the characteristic out-of-plane vibrational mode, $A^1$, located at 120.7 cm$^{-1}$ and in-plane vibrational mode $E^2$, located at 140.5 cm$^{-1}$. Raman spectroscopy of Te deposited on NaCl spin-coated on SiO$_2$/Si substrate confirms the emergence of $A^1$ and $E^2$ modes located at 120.9 cm$^{-1}$ and 140.5 cm$^{-1}$ (**Figure 3a** bottom panel). **Figure 3b** summarizes the intensity variations of $A^1$ and $E^2$ Raman modes for the as-grown Te with and without NaCl. Raman intensity of these two peaks for the as-grown Te on SiO$_2$/Si undergoes a large variation across the substrate, exhibiting non-homogeneous deposition of Te . This can be attributed to the non-uniform accumulation of material at different parts of the as-grown Te on SiO$_2$/Si confirming deposition of nanopillars. On the other hand, Raman peak intensity variation of the as-grown Te on SiO$_2$/Si with NaCl is minimized. This fact can stem from the more uniform deposition of Te in the case of using NaCl as salt additive. To obtain a deeper insight into the morphological properties of the Te deposited on SiO$_2$/Si with and without salt additive, AFM in tapping mode is performed on 5 μm × 5 μm scan area of the samples (**Figure 3c-d**). The average grain size of the samples assessed by the self-correlation methods (**Figure SI-6**) reveals that Te deposited with the spin-coated NaCl has an average grain size of 760 ± 21 nm, while Te deposited on SiO$_2$/Si without NaCl has an average grain size of 478 ± 15 nm. The large difference observed in the grain size can be attributed to the significant effect of the NaCl salt additive on the suppression of nucleation density, thus increasing the lateral size of the grown material [34,35].

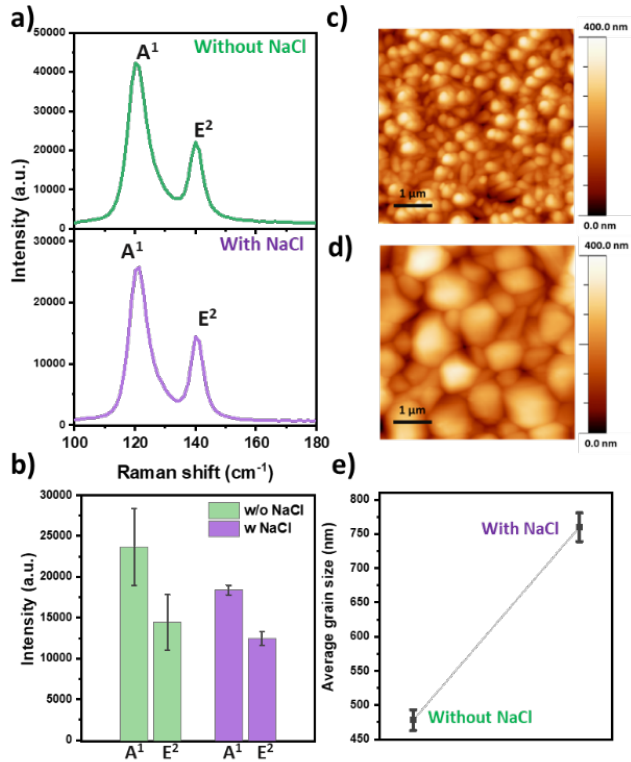

*Figure 3 Raman spectra acquired of the as-grown Te nanostructure on SiO$_2$/Si substrate a) without NaCl on top panel and with NaCl on bottom panel b) comparison of the average Raman intensity of the $A^1$ and $E^2$ peaks acquired on the as-grown Te nanostructure on SiO$_2$/Si substrate without and with NaCl. AFM topography imaging acquired on 5 μm × 5 μm of the as-grown Te nanostructure on SiO$_2$/Si substrate c) without NaCl d) with NaCl e) comparison of the average grain sizes on the as-grown Te nanostructure on SiO$_2$/Si substrate without and with NaCl.*

*1.2. Downscaling tellurium*

After examining the effect of NaCl as salt additive, on the enhancement of lateral growth of Te rather than vertical growth, we aim at downscaling Te by examining the effect of temperature and growth time without the usage of any salt additive and avoiding any chemical contamination. The temperature can have an important role on the thickness of deposited Te on SiO$_2$/Si substrate. In this regard, we used the SiO$_2$/Si substrate on the ceramic boat in the middle of the downstream, without any additive and applying the temperature of 350 ºC with 10 sccm carrier gas flux for 30, 40, and 50 minutes of growth time (**Figure SI-1b**). The AFM tapping phase and topography figures of tellurene flakes grown at 30, 40, and 50 minutes of growth time are summarized in **Figure 4**. The height profile of the AFM images is consistent with a single layer thickness. It shows that tellurene flakes in the form of elongated hexagons are grown. Increasing the growth time on the other hand leads to increasing the thickness from 0.6 nm to 1.2 nm and 1.8 nm. As shown in **Figure SI-7** the maximum lateral length of tellurene semi-hexagonal flakes is increased by the prolonged growth time from 4 μm at 30 minutes to 6 μm at 40 minutes and finally reaching 10 μm at 50 minutes. Furthermore, the distribution of tellurene nanoflakes is enhanced by increasing the growth time as it can be observed in tapping phase images summarized in Figure SI-7 d-f performed on 20 μm × 20 μm scan area. In particular, Figure SI-7 e-f shows the tellurene flakes are grown both in the form of semi-hexagonal and elongated 1D flakes that can act as the seed layer for the formation of semi-hexagonal flakes. Increasing growth time can have an important effect on stabilizing the final structure of the flakes as we observed here its effect on increasing density and lateral size of the flakes.

The Raman spectra obtained from three different positions on the tellurene nanoflakes are depicted in **Figure SI-8**. These spectra reveal a noticeable blueshift, in both the $A^1$ and $E^2$ Raman modes of Te confirming the downscaling of the tellurene thickness compared to the same measured peaks in the previous section. In more detail, the $A^1$ peak of the tellurene nanoflakes experiences a blueshift to the range of 128-130 cm$^{-1}$. Furthermore, the $E^2$ mode in the Raman spectra also demonstrates a blueshift to the range of 144-145 cm$^{-1}$.

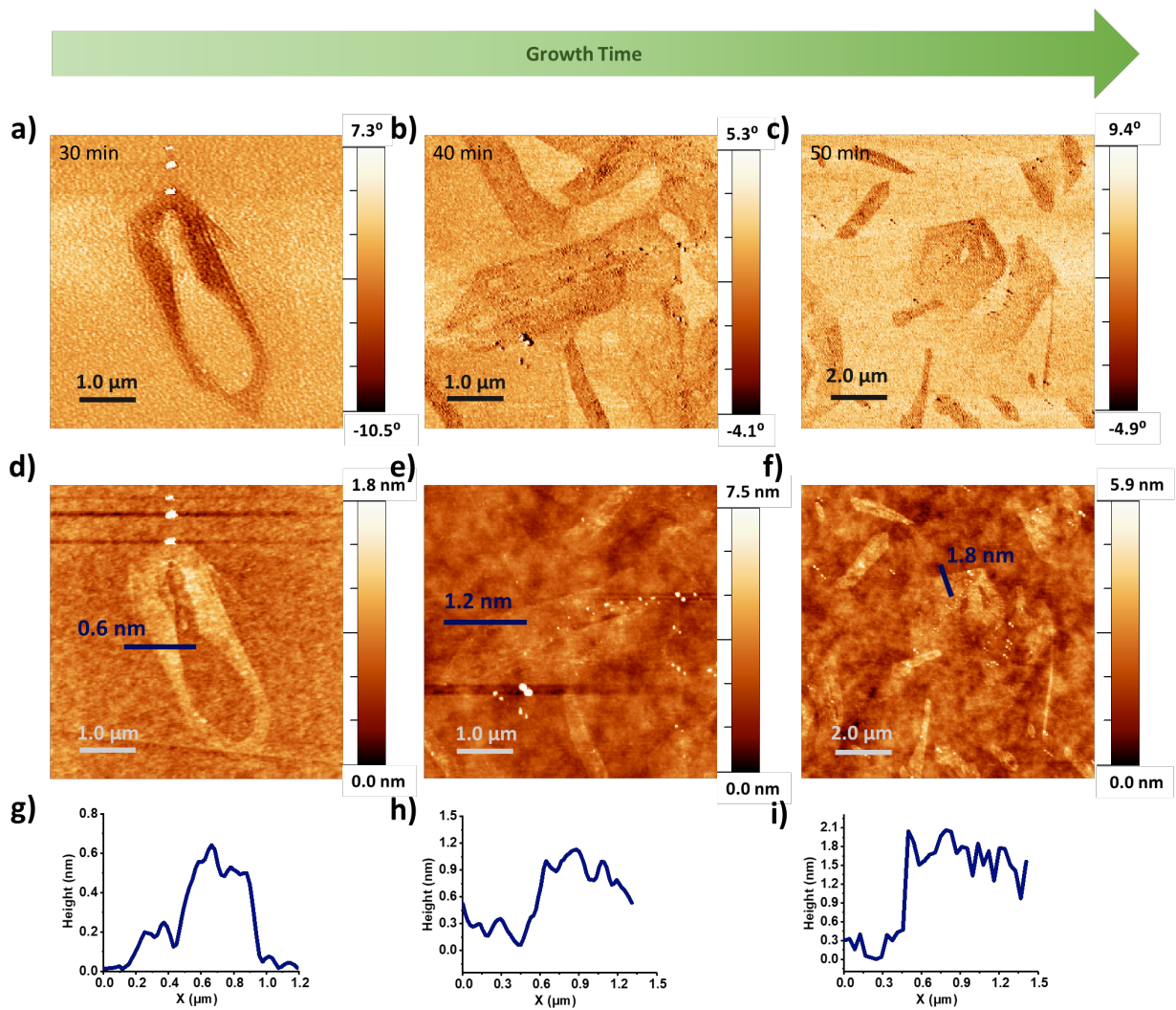

*Figure 4 AFM a) tapping phase d) topography and g) height profile of tellurene flake grown at 30 minutes growth time b) tapping phase e) topography and h) height profile of tellurene flake grown at 40 minutes growth time c) tapping phase f) topography and i) height profile of tellurene flake grown at 50 minutes growth time*

Figure **5a-b** shows topographic and tapping phase imaging of monolayer of tellurene flakes with 0.4 nm thickness. The monolayer flake of tellurene shows a hexagonal structure with a hole in the middle of the flake. To get a deeper understanding of the morphological distribution of tellurene flake, a topography imaging using AFM on 30 μm × 30 μm scan area has been performed. As recorded in **Figure 5d** tellurene flakes can be achieved in hexagonal and elongated crystal shapes. The flake delimited by green dashed line in Figure 5d and its corresponding 3D topographic illustration in **Figure 5e** represent a few-layer hexagonal flake of tellurene. In the middle of the few-layer flake despite the monolayer one there exists a higher thickness cross-like accumulation of Te, which can act as preferential site for the nucleation of Te that during the growth evolve into the next hexagonal layer. The middle area of the flakes is a favorable point for the initiation of growing of subsequent layers and leads to a height distribution histogram of flakes with three distinct peaks at 4, 5.2, and 8.4 nm (**Figure 5e**).

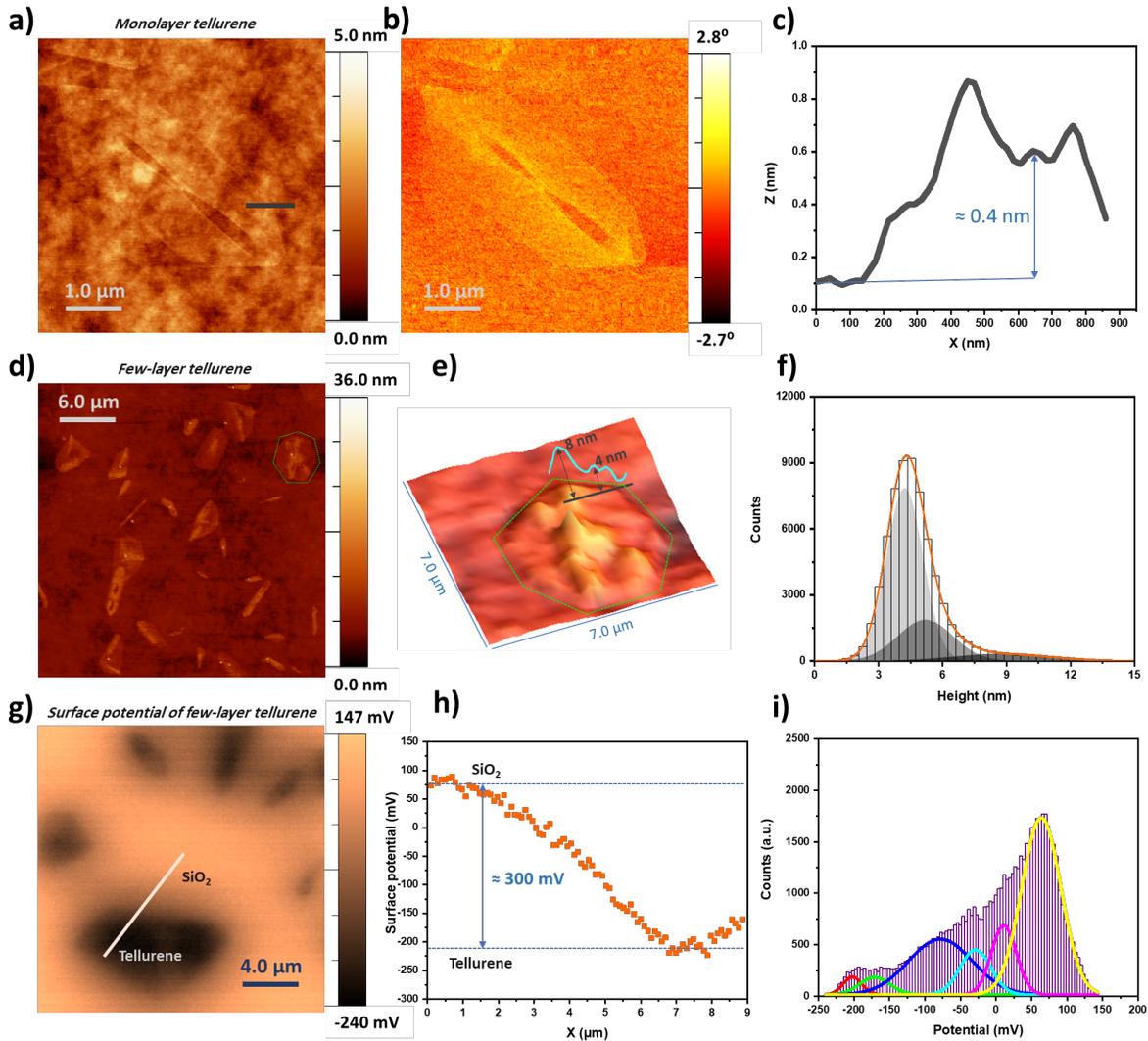

*Figure 5 a-b) AFM topography and tapping phase image of the hexagon tellurene flake c) height profile corresponding to the gray line at the edge of flake in panel a representing monolayer tellurene. d) AFM topography imaging performed on the 30 μm × 30 μm scan area of the few-layer tellurene flake e) 3D topographic scan of the depicted hexagonal flake with dashed green line in panel d illustrating the height variation in the middle of the flake f) histogram distribution of the height of few-layer tellurene flakes in panel d g) surface potential map obtained from KPFM performed on the tellurene flakes representing striking difference of surface potential between tellurene flakes and SiO$_2$ background h) corresponding potential variation along the line in panel g i) histogram distribution of the surface potential map of the panel g.*

To gain insight into the electronic properties of the tellurene flakes at the nanoscale we carried out AFM investigations in kelvin probe configuration. The surface potential variation of the 20 μm × 20 μm scan area is depicted in **Figure 5g** and indicates a striking electrostatic contrast between tellurene flakes and SiO$_2$ background. The tellurene flakes appear as dark pitches and negatively charged respect to the background. The surface potential variation of ≈ 300 mV is observed between tellurene and SiO$_2$ substrate (Figure 5h). Furthermore, the histogram distribution shows multiple peaks at negative potential values of -202, -170, -78, and -29 mV, which correspond to tellurene flakes, and positive potential values associated with the SiO$_2$ surface (**Figure 5i**).

## Conclusions

We demonstrated the capability to control the dimensionality of synthetic Te sheet on the substrate by achieving wafer scalability through the usage of salt-made additives in the growth processing and by reducing their thickness down to the 2D level. On one hand, the as-grown Te film using spin-coated NaCl solution on $SiO_2$/Si substrate yields the formation of compact granular grains with lateral size of $760 \pm 21$ nm, and a pronounced enhancement compared to the as-grown Te nanopillars with diameter of $478 \pm 15$ nm grown on the $SiO_2$/Si substrate without NaCl. On the other hand, without exploiting any promoter, through a single step vapor transport approach we achieve the thinnest reported tellurene hexagonal flakes down to monolayer regime. The lowest thickness of as grown tellurene flakes were 0.4 nm with lateral size of 4 μm.

## Acknowledgements

We acknowledge Mr. Mario Alia and Mr. Simone Cocco (CNR-IMM) for technical support. We acknowledge financial support from ERC-CoG grant n. 772261 "XFab".

**Supporting Information**

The temperature ramp applied to the upstream and downstream furnaces and the carrier gas flux associated with the positional-dependent growth section (Figure SI-1a) and the downscaling tellurium section (Figure SI-1b).

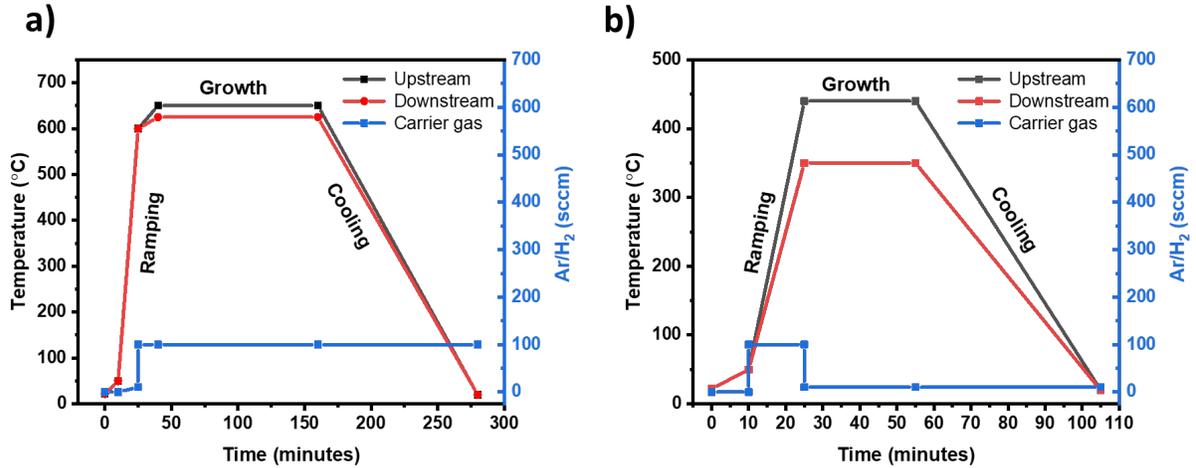

*Figure SI-1 a) Temperature ramp applied to the upstream and downstream furnaces in the first section b) Temperature ramp applied to the upstream and downstream furnaces in the second section*

The diameter distribution of Te nanostructures was determined using ImageJ software, which analyzes SEM images displayed in Figures 1b and 1c. The distribution of the diameter of grown Te pillar-like structures are represented in histogram Figure SI-2. To quantitatively analyze the data, each histogram has been fitted using MATLAB software. We noticed that the Lognormal distribution functional must be used to minimize the chi-squared values of 2.4 and 1.1 of the histogram distributions of Te pillar-like structures of Figures 1b and 1c of the manuscript and displayed figure SI-2. The Lognormal distribution function fitted to the diameter of solid pillar structures of Figure 1b, shows a distribution that is ranged from 0.3 to 0.8 μm with the mean value calculated to be $0.6 \pm 0.1$ μm. For the layered pillar structures associated with Figure 1c the Lognormal distribution has been used to fit the data ranging from 0.7 to 2.1 μm with the mean value of $1.4 \pm 0.3$ μm which is larger than the first case. The observed difference in the diameter distribution can be attributed to the different growth regimes due to the vicinity of substrate 2 to the Te precursor boat, and having a relatively higher temperature than substrate 1, which causes higher deposition rates presumably leading to larger nanostructures. Meanwhile the deposited Te nanostructures on substrate 1 experience a moderate deposition rate due to the larger distance and lower temperature with respect to substrate 2 thereby resulting in a lower average diameter. As a phenomenological conclusion, we can speculate that a control of the diameter size of the pillars can be obtained by tuning the Te source-substrate distance.

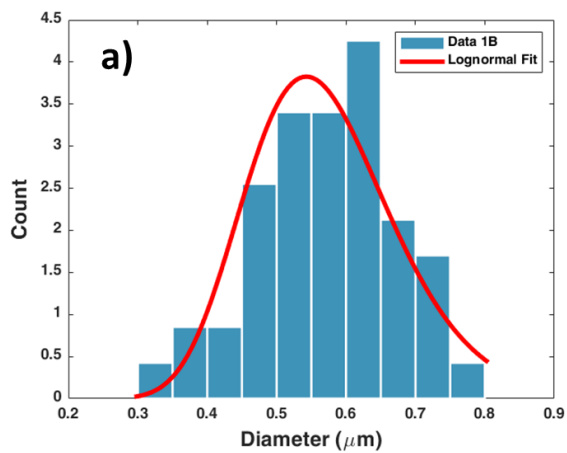 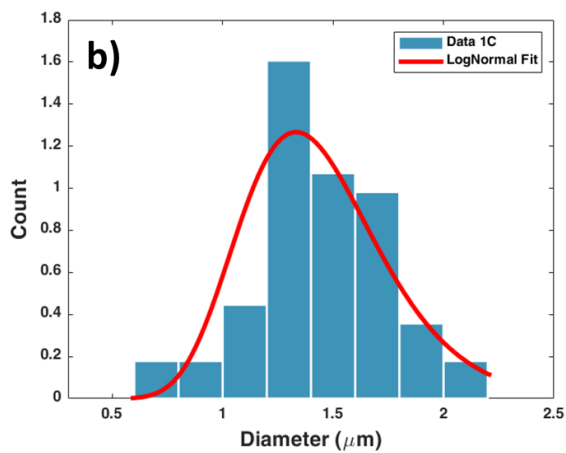

*Figure SI-2 Average diameter distribution of a) As-grown Te pillar-like nanostructures on substrate 1 in Fig. 1b b) As-grown Te pillar-like nanostructures grown on substrate 2 in Fig. 1c*

The tellurium nanostructures obtained in substrates 3 (Figure SI-3 a) have a similarity with the traditional Italian pasta called "pennette" (Figure SI-3 b). As a result, we called these nanostructures "pennette-like" Te nanotubes.

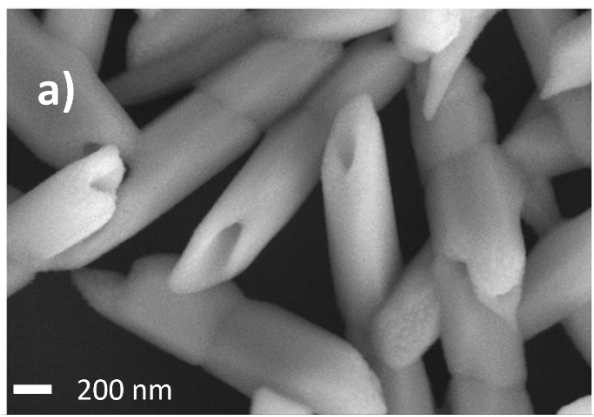 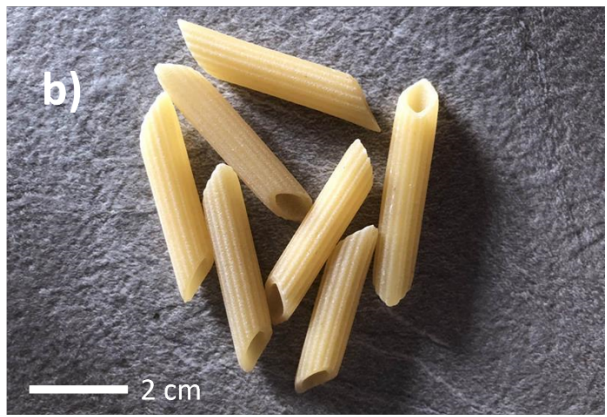

*Figure SI-3 a) High magnification SEM image of the as-grown "pennette-like" Te nanostructures on substrate 3 b) traditional Italian pasta called "pennette"*

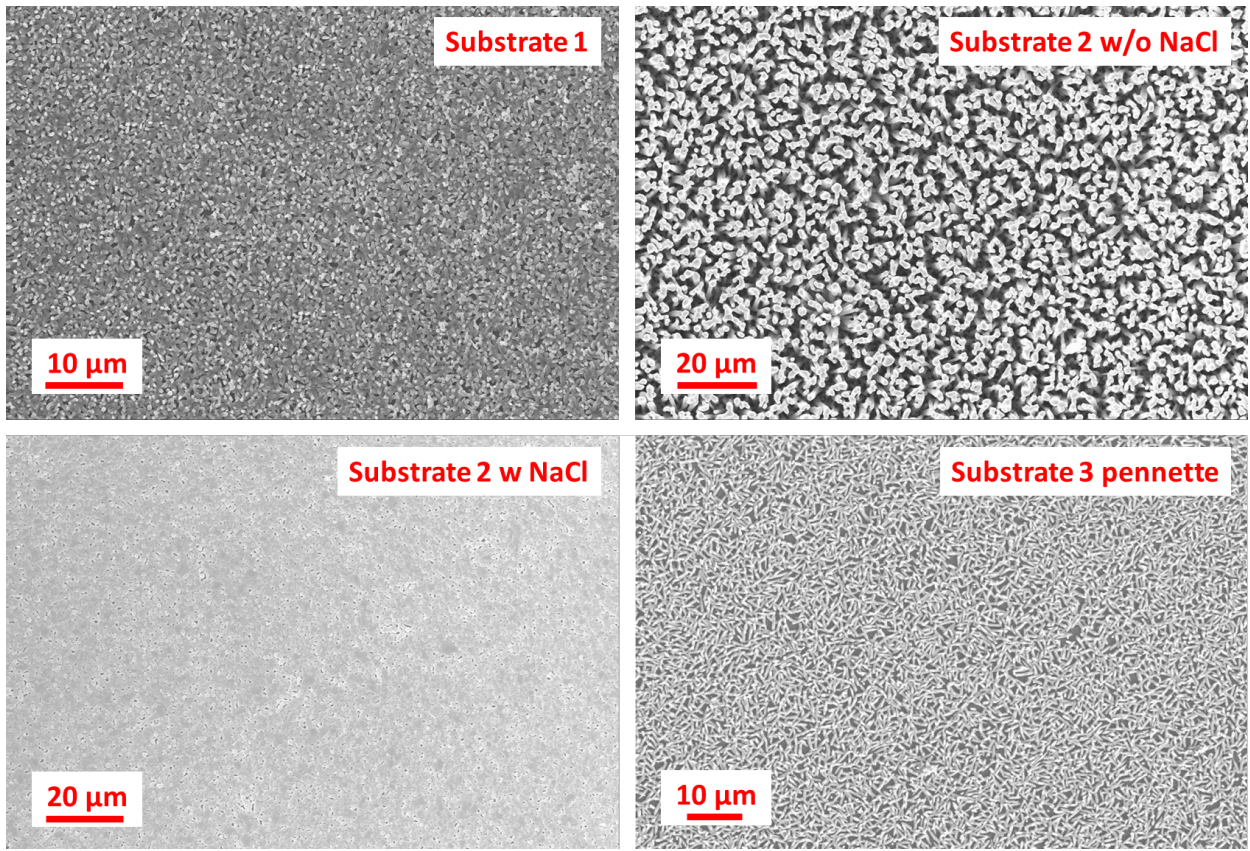

Figure SI-4 Large-scale SEM images of the Te nanostructures grown on substrate 1, substrate 2 without promoter, substrate 2 with promoter and substrate 3.

The Raman spectroscopies can be used as a powerful tool to characterize the homogeneity of the growth at the macro-scale. In Figure SI-5 we map the homogeneity level of substrate 1, substrate 2 without NaCl, substrate 2 with NaCl and substrate 3 by means of positional Raman spectroscopy scanning. We arbitrarily set the origin of the positional scanning in the point (x = 0 mm) of the substrates closer to the tellurium source and we identify the points at $x_1$ to $x_4$ with the distance between two adjacent points of 0.5 cm . The evolution of the Raman spectra as a function of the position distance identifies a clear variation of the Raman peak intensities of the characteristic $A^1$ and $E^2$ modes. We noticed the average percentage of intensity variation for substrate 2 obtained using NaCl promoter is 2.5 %, while for substrate 1, 2 without promoter, and 3 are as high as 47 %, 19 %, and 40 % respectively.

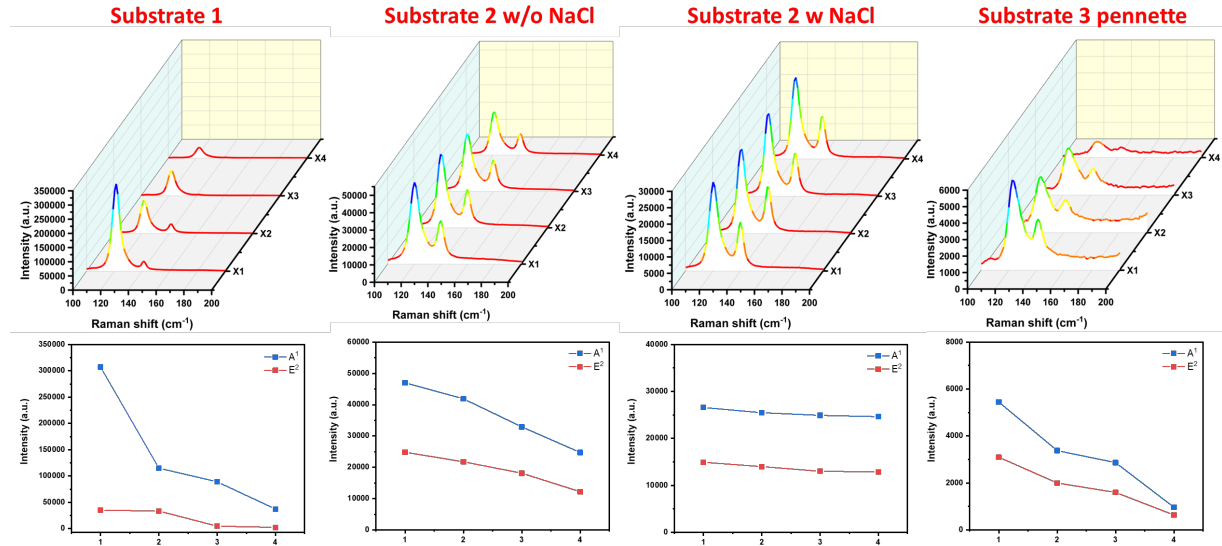

*Figure SI-5 Acquired positional Raman spectra across the Te nanostructures grown on substrate 1, substrate 2 without promoter, substrate 2 with promoter and substrate 3*

The AFM topography imaging on 10 μm × 10 μm scan area of the as-grown Te on bare SiO$_2$/Si (Figure SI-6a) and on NaCl spin-coated on SiO$_2$/Si substrate (Figure SI-6b). The average grain size of the samples assessed by the self-correlation methods reveals that Te deposited on bare SiO$_2$/Si without NaCl has an average grain size of 475 nm (Figure SI-6c), while Te deposited with the spin-coated NaCl has an average grain size of 785 nm (Figure SI-6d).

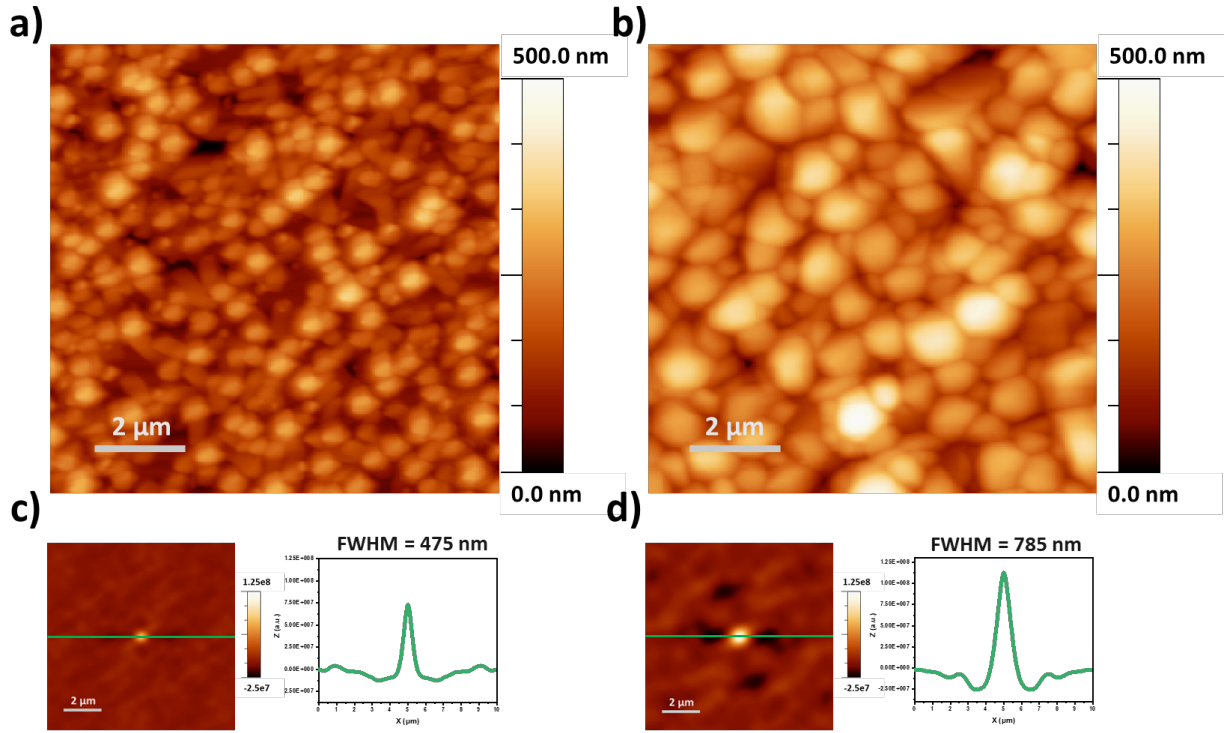

*Figure SI-6 a) AFM topography imaging performed on 10 μm × 10 μm scan area of the Te deposited on SiO$_2$/Si substrate a) without NaCl and b) with NaCl. Self-correlation filter applied to the topography images of c) figure d) figure b with their FWHM of the central peak measuring the average grain size*

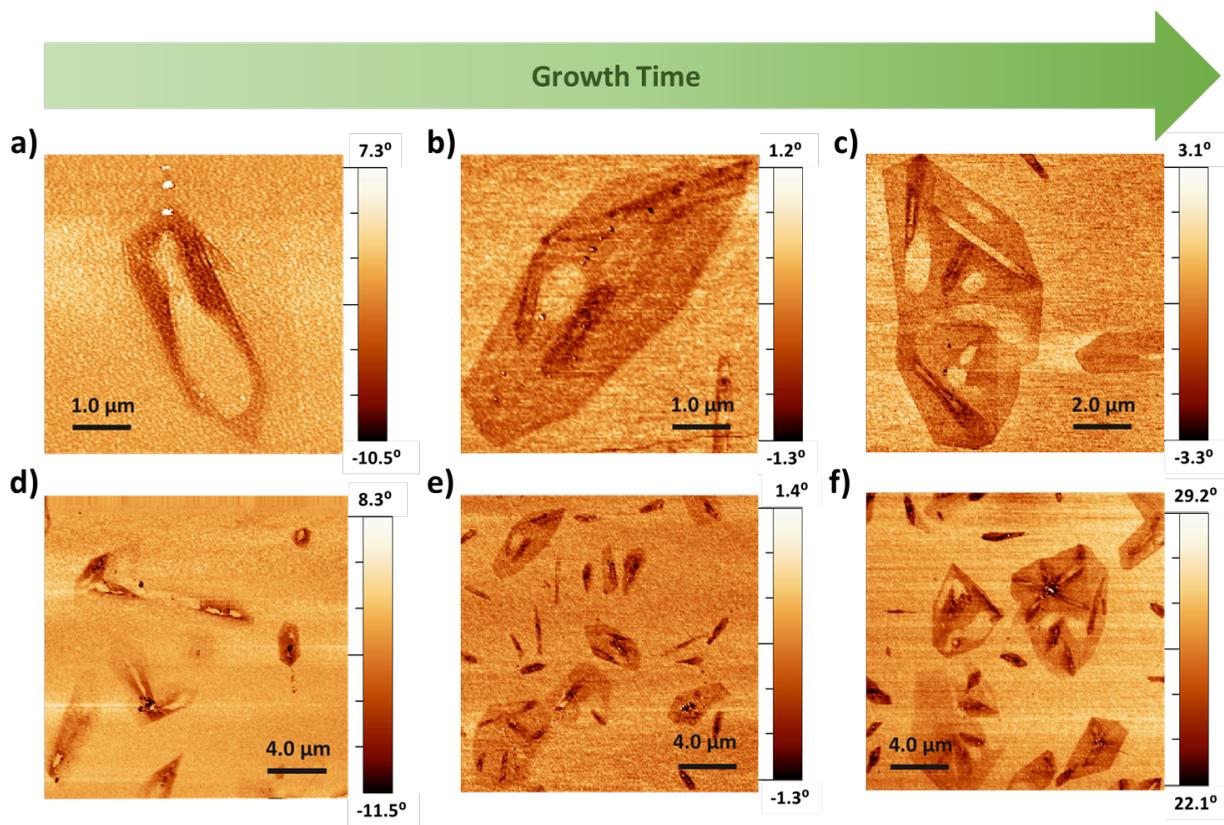

*Figure SI-7 AFM tapping phase images of tellurene flakes grown at a,d) 30 minutes b,e) 40 minutes and c,f) 50 minutes of growth time*

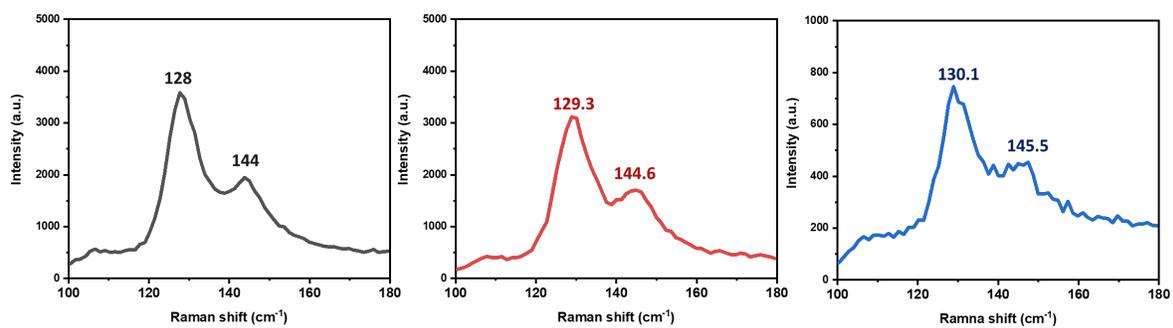

*Figure SI-8 Raman spectra acquired at different points of the as-grown 2D tellurene flakes*